\documentclass[preprint2]{aastex62}

\shorttitle{Alma reveals a collision between protostellar outflows in BHR~71}
\shortauthors{Zapata et al.}

\begin{document}

\title{ALMA reveals a collision between protostellar outflows in BHR~71}

\correspondingauthor{Luis A. Zapata}
\email{l.zapata@irya.unam.mx}

\author{Luis A. Zapata}
\affiliation{Instituto de Radioastronom\'\i a y Astrof\'\i sica, Universidad Nacional Aut\'onoma de M\'exico, P.O. Box 3-72, 58090, Morelia, Michoac\'an, M\'exico}

\author{Manuel Fern\'andez-L\'opez}
\affiliation{Instituto Argentino de Radioastronom\'\i a (CCT-La Plata, CONICET; CICPBA), C.C. No. 5, 1894, Villa Elisa, Buenos Aires, Argentina}

\author{Luis F. Rodr\i\' guez}
\affiliation{Instituto de Radioastronom\'\i a y Astrof\'\i sica, Universidad Nacional Aut\'onoma de M\'exico, P.O. Box 3-72, 58090, Morelia, Michoac\'an, M\'exico}

\author{Guido Garay}
\affiliation{Departamento de Astronom\'\i a, Universidad de Chile, Camino el Observatorio 1515, Santiago, Chile}

\author{Satoko Takahashi}
\affiliation{Joint ALMA Observatory, Alonso de Cordova 3108, Vitacura, Santiago, Chile}

\author{Chin-Fei Lee}
\affiliation{Academia Sinica Institute of Astronomy and Astrophysics, P.O. Box 23-141, Taipei 106, Taiwan}

\author{Antonio Hern\'andez-G\'omez}
\affiliation{Instituto de Radioastronom\'\i a y Astrof\'\i sica, Universidad Nacional Aut\'onoma de M\'exico, P.O. Box 3-72, 58090, Morelia, Michoac\'an, M\'exico}
\affiliation{IRAP, Universit\'e de Toulouse, CNRS, UPS, CNES, Toulouse, France}



\begin{abstract}
For a binary protostellar outflow system in which its members are located
so close to each other (the separation being smaller than the addition of the widths of the flows) 
and with large opening angles, the collision seems unavoidable regardless of the orientation of the outflows.
{This is in contrast to the current observational evidence of just a few regions with indications of colliding outflows.}
Here, using sensitive observations of the Atacama Large Millimeter/Submillimeter Array (ALMA),
we report resolved images of carbon monoxide (CO) towards the binary flows associated with the BHR~71 
protostellar system. These images reveal for the first time solid evidence that their flows are partially colliding, 
increasing the brightness of the CO, the dispersion of the velocities in the interaction zone, and changing part of the orientation 
in one of the flows.  Additionally, this impact opened the possibility of knowing the 3D geometry of the
 system, revealing that one of its components (IRS~2) should be closer to us.    
\end{abstract}



\section{Introduction} \label{sec:intro}
The forces that govern the energy budget of molecular clouds are still a matter of study. Gravity, magnetic fields and turbulence are the main actors of this play, but their relative importance is still unknown and can depend on each particular case. In particular the lifetimes of molecular clouds are a matter of debate since turbulence is expected to decay in such a short time (a crossing time) that the star formation rate should be much larger than observed. The alternatives are that either molecular clouds are stand by large-scale magnetic fields, resupplied with turbulence by internal feedback processes, or are just transient structures.

{In the heart of molecular clouds, the density of stars can be as high as 1000 stars per cubic parsec \citep[like in Orion, ][]{tes1999}, implying a high density of outflows which makes them a natural source of turbulence at a range of different spatial scales from a few AUs to parsec scales \citep{bal1996,bal2001,mat2002,mas2015}. Jets impinge energy and momentum by entraining the gas of a cloud at the boundaries of the outflow \citep{can1991}. An isotropically distribution of jets in a cloud may produce turbulence in a broad range of spatial scales. Moreover, if the volume of the outflows somehow increases, or if they interact with the medium creating more vortices, then the turbulent energy would increase adding support to the cloud against gravity. Therefore, different interactions between outflows and the surrounding medium can enhance cloud's turbulence.}

{Jets and outflows are ubiquitous in star-forming regions and their propagation through the Interstellar Medium has been studied in detail and is reasonably well understood \citep[e.g.,][]{lee2001,bal2007,bal2016}. The interaction between jets and dense parts of clouds or clumps has received also some attention \citep{rag1995}, as well as the interaction with side-winds \citep{can1995} or the interaction of jets of runaway protostars with the quiescent Interstellar Medium \citep{can2008}. Precession effects of a single jet due to the orbital motion of a binary companion or the misalignment of the ejection mechanism with the system rotation axis has also been the subject of several observational and theoretical work. However, interactions between jets or outflows themselves, remain quite unexamined. In spite of being likely occurring in typical star-forming regions \citep{cun2006}, interactions between two or more outflows have not been clearly observed.}

{As we just said,} there is little evidence of colliding protostellar outflows. One of the best cases presented of this phenomenon 
is found in the IR-dark ``tail" that crosses the IC 1396N globule \citep{bel2012}.  CO 
observations show two well collimated protostellar outflows associated with young 
stars at the northern part of the globule, these outflows are located close to the plane of the sky and probably are 
colliding (the blue-shifted lobes) with each other toward the position of a strong 2.12 $\mu$m H$_2$ line emission feature. 
This H$_2$ feature is recognized as being excited by the impact of the two protostellar outflows.  A less clear case of colliding
outflows is the one found in the massive infrared dark cloud clump, G028.37+00.07-C1 \citep{kon2017}.  In that case,
the collimated outflows emanating from the massive protostellar objects C1-Sb and C1-Sa seem to collide, but  it is not
clear if this involves a real physical interaction or is simply a projection effect.
 
There is also some evidence of outflow collision but with dense parts of their natal molecular clouds \citep{raga2002,cho2005}.   
The interaction between the protostellar outflow and the dense cloud causes an abrupt deflection on the orientation of the flow. 
This effect has been observed in both cases, the HH210 and NGC 1333 IRAS 4A molecular flows \citep{raga2002,cho2005}. 

The bipolar molecular outflows emanating from BHR~71 are powered by two young stars separated by 15$''$ or 3500 AU (at an assumed distance of 200 pc)
called IRS~1 and IRS~2 \citep{bou1997}.  IRS~1 is classified as a Class I/0 protostar with a bolometric luminosity of 13.5$\pm$1.0 L$_\odot$, a dust temperature
of 25 K, and a mass of the dusty envelope of 2.1$\pm$0.4M$_\odot$, while IRS~2 (which is classified as a Class 0 protostar) has a       
bolometric luminosity of 0.5$\pm$0.1 L$_\odot$, a dust temperature of 26 K, and a mass of the dusty envelope of 0.05$\pm$0.02 M$_\odot$ \citep{che2008}.
More recently, using Herschel observations, it has been estimated for BHR~71 (which includes binary)  an envelope mass 
of 19 M$_\odot$ inside a radius of 0.3 pc and a central luminosity of 18.8 L$_\odot$ \citep{yan2017}. 
The outflow from IRS 1 is larger in extent, more massive, and dominates the CO emission \citep{gar1998,bou2001}. 
The IRS~2 bipolar outflow was revealed by the APEX telescope
using  the $^{12}$CO(4-3) line given its better angular resolution and sensitivity compared with other single dish telescopes \citep{par2006}.  
Contrary to the outflow from IRS~1 its blueshifted emission is found to the northwest, while its redshifted emission is found to the southeast \citep{par2006}.  
The temperature in the outflow lobes from the two sources is very different. For IRS~2 the temperature ranges between 30 and 50K, while for IRS~1
 it rises up to 300K. This again suggests the dominance of the outflow from IRS~1 as observed in other studies \citep{par2006,yil2015,gus2015,ben2017}.

\begin{figure*}
\centering
\includegraphics[angle=0, scale=0.65]{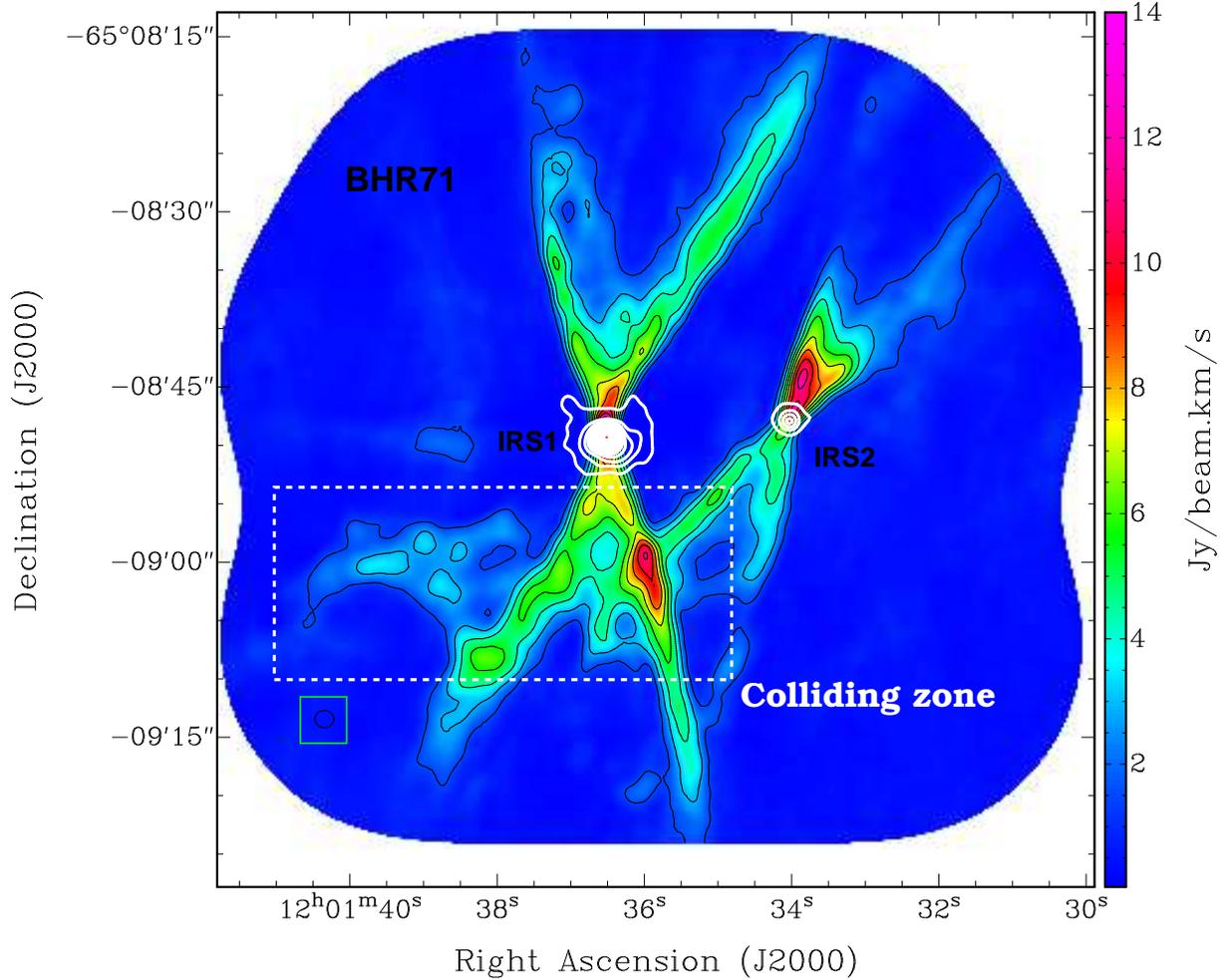}
\caption{ALMA $^{12}$CO(2-1) integrated intensity (moment zero) color-scale/contour (black), and 1.3 mm
continuum (white contours) images of BHR~71. The half-power synthesized beam size is shown in the bottom left corner. 
The peak intensity scale-bar is shown in the right. The black contours are starting
from 10$\%$ to 80$\%$ in steps of 10$\%$ of the intensity peak. The CO intensity peak is 13.9 Jy Beam$^{-1}$ Km s$^{-1}$. 
The white contours are starting from 2$\%$ to 90$\%$ in steps of 2$\%$ of the intensity peak. 
The 1.3 mm continuum intensity peak is 0.4 Jy Beam$^{-1}$. Please note how in the colliding zone the outflow intensity is increased
by a factor of two or three the average values within the flow.
 } 
\end{figure*}

\section{Observations} \label{sec:obs}

The archive observations of the BHR~71 protostellar binary system were carried out with ALMA 
at Band 6 in 2015 Jan 17 as part of the Cycle 2 program 2013.1.00518.S. 
The observations used 35 antennas with a diameter of 12 m, yielding baselines with projected lengths 
from 15 to 348.5~m (11--268~k$\lambda$).   The primary beam at this frequency has a full width at half-maximum 
of about 27$''$. In order to cover the central part of the outflows emanating from BHR~71, it was made a small mosaic of fourteen centers 
distributed in a Nyquist-sampled grid.  The integration time on-source was 14.1 minutes, approximately 1 min. for each pointing 
in the mosaic.  The continuum image was obtained averaging line-free channels
from six spectral windows (of 0.059 and 2.000 GHz width) centered at rest frequencies: 219.563 GHz (spw0),  220.402 GHz (spw1), 
217.998 GHz (spw2), 230.548 GHz (spw3), 231.332 GHz (spw4),  and 232.332 GHz (spw5) which covers a total bandwidth of 4.236 GHz. 
These windows were centered to observe different molecular species as the $^{13}$CO(2-1), C$^{18}$O(2-1), and $^{12}$CO(2-1).
In this study, we concentrate in the $^{12}$CO(2-1), which is well known to be an excellent outflow tracer.  This line was centered in the 
spectral window spw 3 (at 230 GHz) and has a channel spacing of 0.08 km s$^{-1}$.  Strong CO emission is detected, see Figure 1.

\begin{figure*}
\centering
\includegraphics[angle=0, scale=0.65]{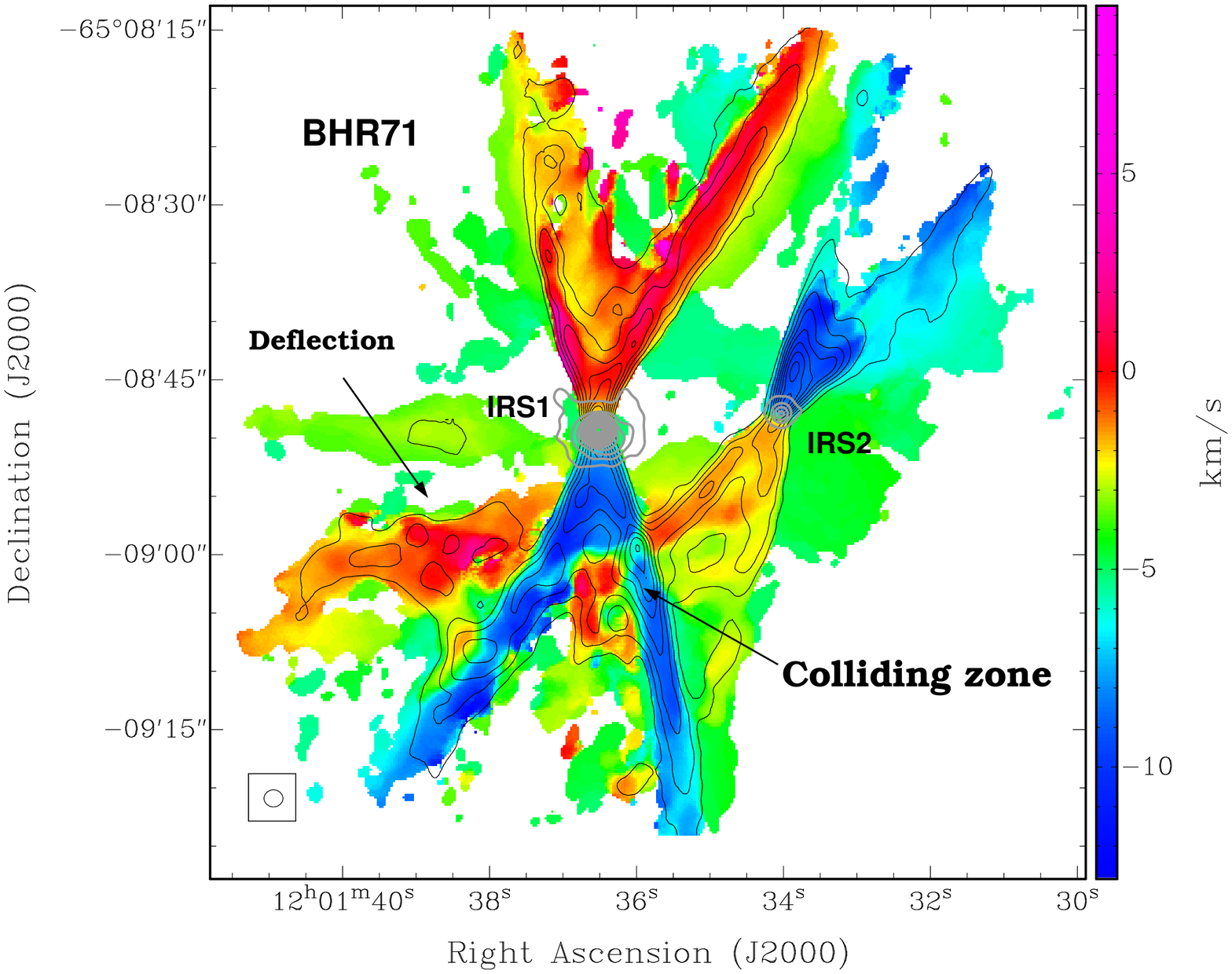}
\caption{ALMA $^{12}$CO(2-1) intensity-weighted velocity (moment one) (color-scale), $^{12}$CO(2-1) integrated intensity (black contours), 
and 1.3 mm continuum (grey contours) images of BHR~71. The half-power synthesized beam size is shown in the bottom left corner. 
The radial velocity scale-bar is shown in the right. The black contours are starting
from 10$\%$ to 80$\%$ in steps of 10$\%$ of the intensity peak. The CO intensity peak is 13.9 Jy Beam$^{-1}$ Km s$^{-1}$. 
The grey contours are starting from 2$\%$ to 90$\%$ in steps of 2$\%$ of the intensity peak. 
The 1.3 mm continuum intensity peak is 0.4 Jy Beam$^{-1}$. 
 } 
\end{figure*}

The weather conditions were reasonably good for this frequency and stable with an average precipitable water vapor of about 3.4 mm 
and an average system temperature of 125 K. The ALMA calibration included simultaneous observations of the 
183 GHz water line with water vapor radiometers, used to reduce the atmospheric phase noise. 
Quasars J1107$-$4449,  J1229$-$6003 and J1229$-$6003 were used to calibrate the bandpass, the atmosphere 
and the gain fluctuations, respectively. Ganymedes was used for the flux amplitude. 
The data were calibrated, imaged, and analyzed using the Common Astronomy Software Applications  CASA \citep{mac2007}.
Imaging of the calibrated visibilities was done using the task CLEAN.  The resulting image {\it rms} noise for the continuum was 
0.4 mJy beam$^{-1}$ at an angular 
resolution of $1\rlap.{''}2 \times 1\rlap.{''}1$ with a PA = $-$63.6$^\circ$.  The ALMA theoretical {\it rms} noise for this configuration, 
integration time, and frequency is about 0.3 mJy beam$^{-1}$, which is very close to the value we obtain in the continuum images.
We used the ROBUST parameter of CLEAN in CASA set to 0. 
For the line emission,  the resulting image {\it rms} noise was 50 mJy beam$^{-1}$  km s$^{-1}$ at an angular 
resolution of $1\rlap.{''}6 \times 1\rlap.{''}4$ with a PA = $-$77.2$^\circ$, and a bandwidth of 0.084 km s$^{-1}$.
A mild outer taper (140~k$\lambda$) was applied to avoid the sparsity of the longest baselines. 
The ALMA theoretical {\it rms} noise for this configuration, 
integration time, and frequency is about 40 mJy beam$^{-1}$ km s$^{-1}$, which is again very close to the value we obtain in the line images.
The resulting 1.3 mm continuum and $^{12}$CO(2-1) line emission images are presented in Figure 1.
We did self-calibration in phase and amplitude using the resulting mm continuum image as model. We then applied the solutions for self-calibration
in the continuum to the spectra.

\section{Results} \label{sec:res}

Figure 1 shows the moment zero or integrated intensity map of the $^{12}$CO(2-1) line emission (CO from now on, unless otherwise indicated) overlaid with the 1.3 mm continuum emission
as revealed by the ALMA observations. The 1.3 mm continuum emission traces the dust {from the envelopes and protoplanetary disks} surrounding IRS~1 and IRS~2.  
For the case of IRS~1, the ALMA observations reveal a small cavity present {in the northern part of its dusty envelope}, likely carved by the redshifted part  
of its bipolar outflow.  From the observations, IRS~1 has a deconvolved size of 0.78$''$ $\pm$ $0.037''$ $\times$ 0.62$''$ $\pm$ $0.039''$ with a Position Angle\footnote{Measured as usual from North to East.} (PA)
of 94$^\circ$ $\pm$ 15$^\circ$ and a flux density of 547 $\pm$ 10 mJy. For IRS~2, we estimated a deconvolved size of 0.89$''$ $\pm$ $0.094''$ 
$\times$ 0.84$''$ $\pm$ $0.099''$ with a PA of 120$^\circ$ $\pm$ 70$^\circ$ and a flux density of 44 $\pm$ 5 mJy. The corresponding physical sizes of 
these deconvolved values are between 120 to 180~AU, which are typical for accreting disks. Assuming that the dust emission is optically thin and isothermal, 
the dust mass (M$_d$) is directly proportional to the flux density (S$_\nu$) as:

$$
M_d=\frac{D^2 S_\nu}{\kappa_\nu B_\nu(T_d)},
$$    

\noindent
where $D$ is the distance to the object, $\kappa_\nu$ the dust
mass opacity, and B$_\nu(T_d)$ the Planck function for the dust
temperature T$_d$.  Assuming a dust mass opacity ($\kappa_\nu$) of 0.015
cm$^2$ g$^{-1}$ (taking a dust-to-gas ratio of 100) appropriated for these wavelengths (1.3 mm) 
for coagulated dust particles with no ice mantles \citep{Oss1994}, 
a typical opacity power-law index $\beta$ = 1.5,
as well as a characteristic dust temperature (T$_d$) of 25 K, we
estimated a lower limit ({provided that} the emission is not optically thin) for the mass of the {most compact part (i.e., not interferometrically filtered) of the disk and envelope system associated with} IRS~1 
of about 0.215$\pm$0.004 M$_\odot$ and {of 0.017$\pm$0.002 M$_\odot$ of the most compact part of the disk and envelope system associated with IRS~2.} 
Note that the level of uncertainty in the {given lower limits are only statistical, and thus proportional to the flux density estimates. The actual} uncertainty could be very large (a factor of two or three) 
given the uncertainty in the opacities. {Comparing the estimated masses to the masses given in \cite{che2008} it seems that a large fraction of the more extended envelope mass is filtered by the ALMA observations, thus revealing the youth of the two protostars that have just a fraction of their surrounding mass into more compact structures (i.e. disks). A mass ratio of the order of M$_{star}$/M$_{disk}$ $\sim$ 10 is frequently found in sources where the masses of the disks and the stars have been estimated \citep{gui1998,rod1998}.} This implies  that the central sources in IRS~1 and IRS~2 are roughly of intermediate- and low- mass, respectively.    

\begin{figure*}
\centering
\includegraphics[angle=0, scale=0.65]{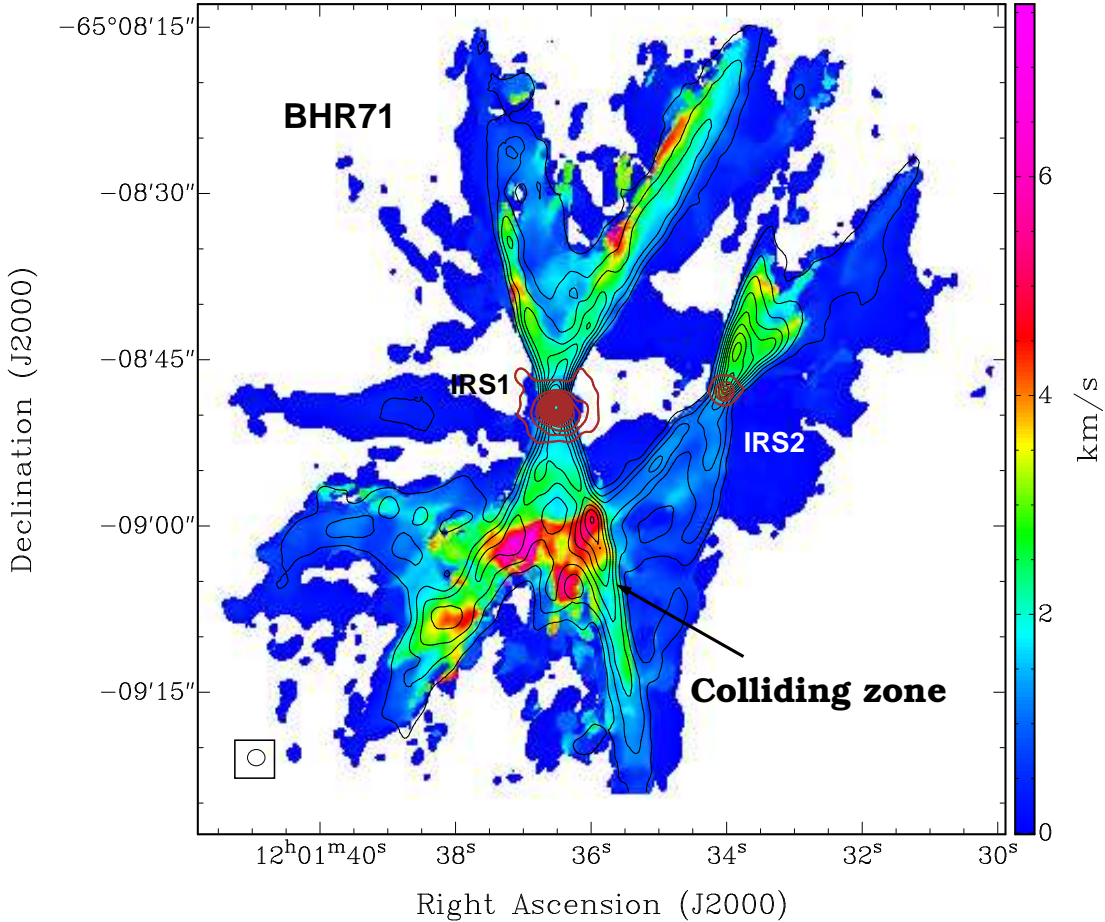}
\caption{ALMA  $^{12}$CO(2-1) intensity-weighted square root velocity (moment two) (color-scale), $^{12}$CO(2-1) integrated intensity (black contours), 
and 1.3 mm continuum (brown contours) images of BHR~71. The half-power synthesized beam size is shown in the bottom left corner. 
The radial velocity scale-bar is shown in the right. The black contours are starting
from 10$\%$ to 80$\%$ in steps of 10$\%$ of the intensity peak. The CO intensity peak is 13.9 Jy Beam$^{-1}$ Km s$^{-1}$. 
The brown contours are starting from 2$\%$ to 90$\%$ in steps of 2$\%$ of the intensity peak. 
The 1.3 mm continuum intensity peak is 0.4 Jy Beam$^{-1}$. 
 } 
\end{figure*}

The CO emission reveals the innermost parts of the two molecular outflows already reported to be associated with BHR~71. A wide-angle structure in both bipolar
outflows is exhibited, with the outflow from IRS~1 being more widely open {(open angle of about $44\degr$, versus $29\degr$ for the outflow of IRS~2)}. These wide-angle structures are likely created by the molecular material being entrained by a collimated jet-like as observed in many other outflows \citep[e.g., ][]{zap2014,zap2015}.  
In this contribution we focus on the study of the morphology and kinematics of the outflows. The mass, energy and momentum will be analyzed elsewhere in more detail.

\begin{figure*}
\centering
\includegraphics[angle=0, scale=0.65]{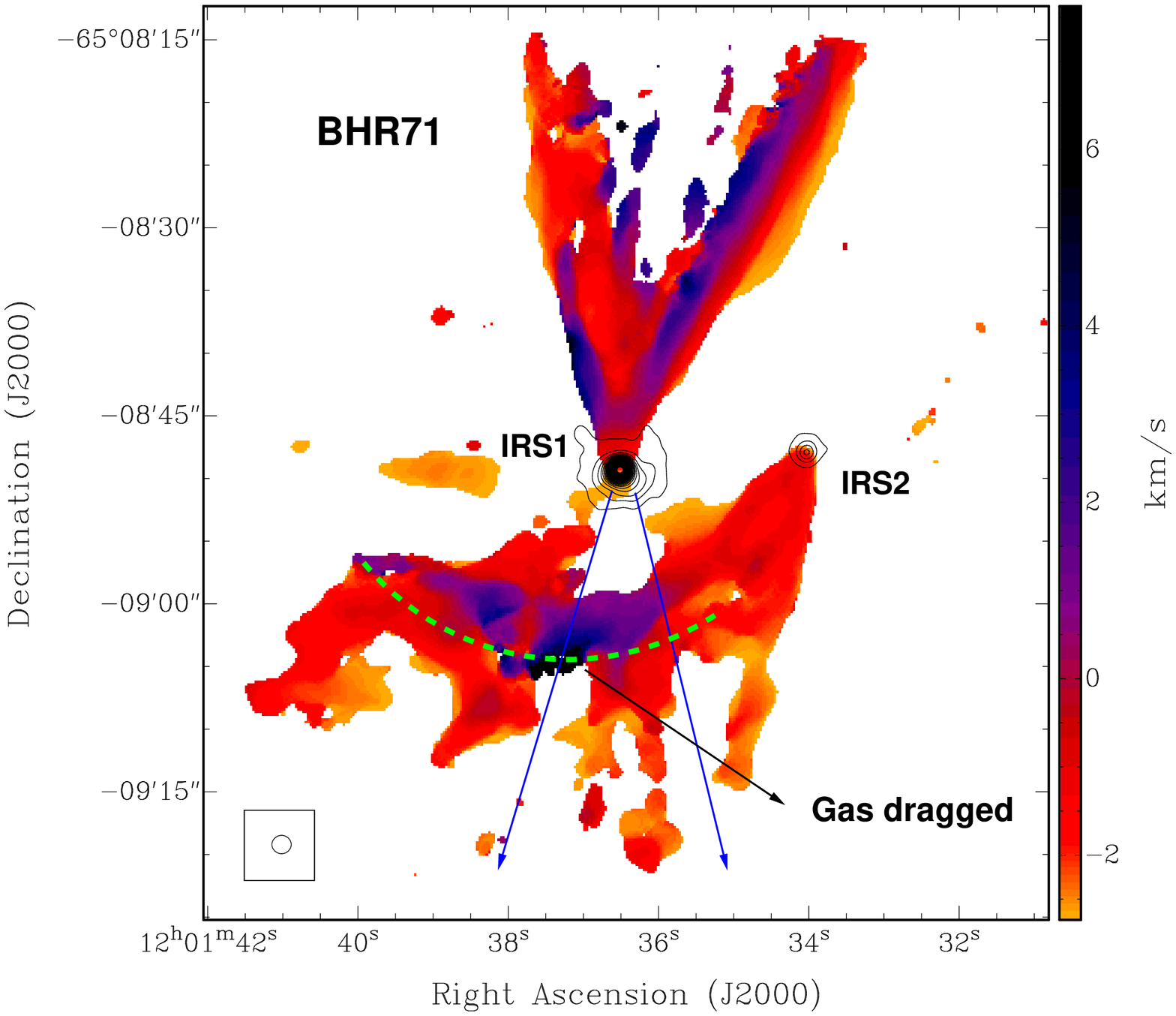}
\caption{ ALMA redshifted $^{12}$CO(2-1) intensity-weighted velocity (moment one) (color-scale) 
and 1.3 mm continuum (black contours) images of BHR~71. The half-power synthesized beam size is shown in the bottom left corner. 
The radial velocity scale-bar is shown in the right.  The black contours are starting from 2$\%$ to 90$\%$ in steps of 2$\%$ of the intensity peak. 
The 1.3 mm continuum intensity peak is 0.4 Jy Beam$^{-1}$.  The blue arrows trace the position of the blue-shifted side of the flow from IRS~1.} 
\end{figure*}  
   
It is easy to appreciate in Figure 1 that in the southwest side of the bipolar outflow from IRS~1 there is a strong and abrupt increase of CO flux of about a factor of two or three the average values within this side of the flow (we have marked this zone in the figure as). This position coincides very well with the intersection in the
plane of the sky between the southwest cavity wall from the IRS~1 outflow and the southeast cavity wall of the outflow from IRS~2.
One can also see that the continuation {of this cavity wall (southeast of IRS~2's outflow) along a} PA of about $133^{\circ}$ bifurcates and seems to be also re-directed into a PA between $70\degr$ and 90$^\circ$. {The southwest cavity wall of the outflow from IRS~2 continues with the same orientation (PA$= 133\degr$) for a longer distance and crosses the IRS~1 outflow cavity close to the southern edge of the ALMA image, with a lower flux enhancement which is washed out in the momentum image due to the frequency averaging. Low angular-resolution $^{12}$CO(3-2) observations have shown that IRS~2's outflow continues apparently unperturbed due southwest \citep{par2006}, but the field of view of the present ALMA CO observations doesn't show this clearly.}

These results {may suggest that both outflows are colliding or just overlapping in projection. In order to further interpret the data, we have made images for the moment 1 (intensity weighted velocity, Figure 2) and moment 2 (velocity dispersion, Figure 3) of the CO emission.}    
 
{Figure 2 presents the moment 1 of the molecular gas of both bipolar outflows. The range of velocities where we integrated to construct this image is from $-$25 to $+$20 km s$^{-1}$. The red-shifted side of the outflow from IRS~1 is located to the north, while the blue-shifted side is located to the south. This is contrary to the bipolar outflow from IRS~2, which has its blue-shifted component toward the northwest, and its red-shifted component to the southeast.  This has been already observed in the APEX observations \citep{par2006}. Figure 2 clearly reveals the presence of red-shifted gas inside the otherwise blue-shifted cavity of the IRS~1 outflow, and also the abrupt change of orientation in the southeast side of part of the IRS~2 outflow. The redshifted gas from the outflow emerging from IRS~2 seems to get redder after crossing the southwest edge of the IRS~1 outflow, probably due to an acceleration process. The morphology and the east-west velocity gradient of this spatially coherent stream of CO emission can be better seen in Figure 4, where we have only integrated toward redshifted velocities from 0~km~s$^{-1}$ to $+$20~km~s$^{-1}$. The wide angle and quite disorganized stream connects the southern lobe of the IRS~2 outflow with the more redshifted emission crossing the southern lobe of the IRS~1 outflow (labeled \textit{colliding zone} in Figures 2 and 3) and the gas due west of the IRS~1 outflow with PA between $70\degr-90\degr$ (labeled \textit{deflection} in Figure 2).}


Moreover, Figure 3 shows a velocity dispersion image of the CO emission. Relatively broad-line emission is generally detected toward the cavity walls of the IRS~1 outflow and the northern lobe of the IRS~2 outflow (about 3~km~s$^{-1}$ on average), where the gas dragged by the jets is entraining the quiescent molecular cloud. Interestingly the cavity walls of the southern lobe of IRS~2 do not present such broad line-widths (about 1~km~s$^{-1}$), which is not surprising, since many outflows present fainter redshifted lobes. However, the most characteristic feature in Figure 3 is the very broad-lines (up to 7.5~km~s$^{-1}$ wide) found along the path of the CO stream crossing the outflow cavity of IRS~1 (i.e.,\textit{colliding zone}).


\section{Discussion}
After inspecting the ALMA CO data, we found strong evidence supporting the existence of an acceleration process along the stream of gas crossing the southern lobe of the IRS~1 outflow. Since this process is spatially coincident with the position where the two outflows of the region overlap, we hypothesize that they may be interacting physically, indeed. To strengthen this hypothesis we summarize here the main observational facts in line with it: (1) the emission from both CO outflows spatially overlap; (2) the stream of gas crossing the outflow cavity from IRS~1 appears as an extension of the southeast cavity wall from the IRS~2 outflow, as if it was deflected; (3) this stream has a velocity gradient to redder velocities from west to east; (4) there is an increase of the CO emission toward the collision zone as expected in region with strong shocks; (5) the collision zone shows much broader lines than the rest of the outflow, matching the expectation for an episode of outflow deflection; (6) there is redshifted gas south of the collision zone that may be dragged due south by the IRS~1 outflow.

\begin{figure*}
\centering
\includegraphics[angle=0, scale=0.4]{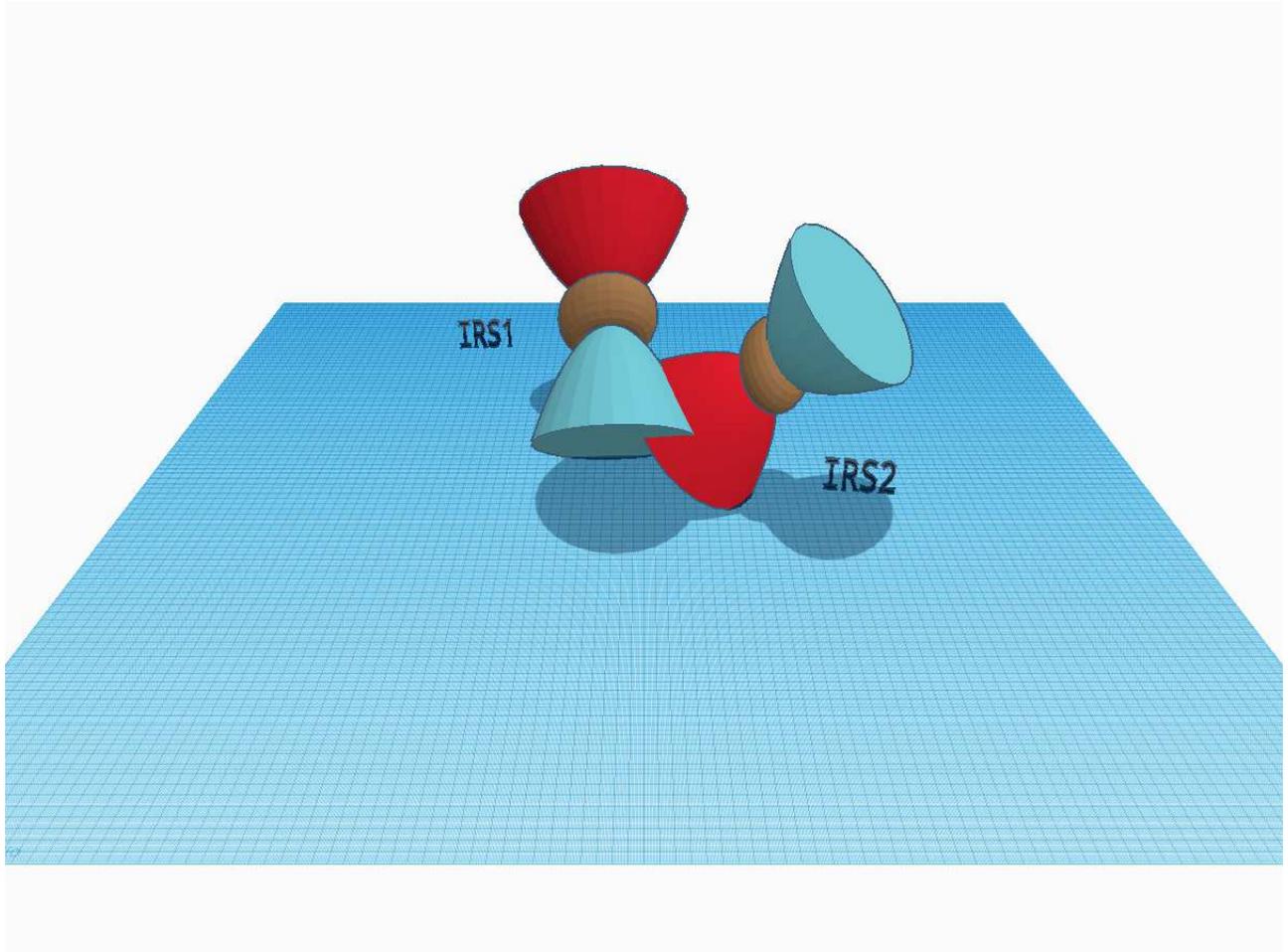}
\caption{3D artistic recreation of the position of the outflows and the young stellar objects in BHR~71. The young stellar object BHR~71-IRS~1 is behind to the position of BHR~71-IRS~2.} 
\end{figure*}

All of these facts are well explained in a collision scenario. However, this explanation needs the outflow from IRS~1 to be more powerful and energetic than the outflow from IRS~2, so that the deflected outflow is that from IRS~2. Given that the IRS~2 protostellar system is less massive and has a lower luminosity, this requirement appears reasonable for now and will be studied in a further contribution elsewhere. Another evidence of the power of the outflow from IRS~1 comes with the optical and IR images that show it has already exit the natal cloud, digging its way out just south of the colliding zone. 

The collision scenario has to explain also why the outflow from IRS~2 runs beyond the cavity of the outflow of IRS~1, apparently unperturbed, and roughly following its original trajectory \citep[as seen e.g., in low-angular resolution $^{12}$CO(3-2) observations in ][]{par2006}. There is also traces in the ALMA CO emission confirming that the outflow from IRS~2 extends further than the colliding zone. All of this indicates that the two flows are only partially colliding. That is, part of the outflow from IRS~2 is kept intact and therefore it proceeds with its original trajectory, while another part physically collides with IRS~1's outflow southern lobe, producing a strong shock and undergoing an accelerated deflection due east. Interestingly, and as a further indication of the strength of the IRS~1 outflow, part of the redshifted emission detected with ALMA appears due south from the collision zone. This emission may be stem from material from the IRS~2 outflow being dragged south by the more energetic IRS1's outflow.

Hence, assuming the collision scenario, a particular geometry is required for the protostars, since the redshifted lobe of the IRS~2 outflow may be impacting the blueshifted lobe of the IRS~1 outflow. The most probable layout is with IRS~2 closer to the observer than IRS~1, as shown in the sketch of Figure 5.  
 
There may be alternative views to the physical interaction between the two outflows. Here we mention that the redshifted emission at the outflows crossing might be caused by the exit of the IRS~1 outflow out of its natal cloud, although this hypothesis cannot explain some of the observational facts -such as the increase on the CO emission (why only in the west side?), or the coherent CO redshifted stream emitting far beyond the cavity walls- and do not accomplish with all the expectations that can be derived from it (e.g., there is no shocked blueshifted emission surrounding the hole dig by the outflow). 

The impacting outflows picture matches all the observations and accomplish with all the expected features breaking as an adequate hypothesis. Were the outflows really colliding, this is one of the few examples to date of this sort of phenomenon and the one with the best evidence gathered (not only spatial and with the adequate resolution but also kinematical). Furthermore, although the collision between two outflows may not be surprising given the probability estimate by \cite{cun2006}, only three cases have been reported so far. In the Appendix, we make a statistical estimate based on the geometry of the collision of two outflows. Given the size of the outflows we let the problem have the separation between protostars as a variable. From our study, we conclude that impacting outflows may be more common than normally thought (as derived from the scarcity of the observational proof) and therefore they may be taken into account when estimating the energetic budget of the molecular clouds.

\section{Conclusions}
\begin{itemize}
\item The sensitivity ALMA CO observations reveal that the outflows in the BHR~71 region are {probably} colliding as indicated for an increasing of the brightness of the CO emission, in the dispersion of the velocities in the impacting zone, and a change in the orientation in one of the outflows (IRS~2). More ALMA observations toward other outflows could reveal more cases of collision, and reveal that this phenomenon occurs more frequent as we discuss in the Appendix. 
   
\item {Provided that} the outflows are colliding in BHR~71, a certain geometry for the system {is implied}. 
One can see how the redshifted lobe of IRS~2 is impacting the blueshifted lobe from the outflow ejected from IRS~1 from Figure 2, and {then derive that IRS~2 should be closer to us than IRS~1}. 
\end{itemize}

\acknowledgments
This paper makes use of the following ALMA data:
ADS/JAO.ALMA\#2013.1.00518.S.  ALMA is a partnership of
ESO (representing its member states), NSF (USA) and NINS (Japan), together
with NRC (Canada), NSC and ASIAA (Taiwan), and KASI (Republic of Korea), in
cooperation with the Republic of Chile.  The Joint ALMA Observatory is
operated by ESO, AUI/NRAO and NAOJ.  LAZ acknowledge financial
support from DGAPA, UNAM, and CONACyT, M\'exico. GG acknowledge financial
support from the CONICyT Project PFB06.

\appendix

\section{The probability of Colliding Flows}

How likely it is that two nearby flows collide? In this appendix we present a simplified discussion of
this topic. We assume that the flows can be represented by circular right capsules
of height $2h$ and radius $r$ (see Figure 6a). The centers of these capsules are separated by a distance $d$.
The axis of the first capsule is given by

$$x_1 = a_1 (h-r) \delta_1; ~~y_1 = b_1 (h-r) \delta_1; ~~z_1 = c_1 (h-r) \delta_1,$$

\noindent where $a_1, b_1, c_1$ are positive numbers between 0 and 1  generated randomly and normalized such that

$$a_1^2 + b_1^2 + c_1^2 = 1.$$

\noindent $\delta_1$ is a discrete variable that goes from $-1$ to $+1$ in steps of 0.01.

The second capsule is given by

$$x_2 = a_2 (h-r) \delta_2 + d; ~~y_2 = b_2 (h-r) \delta_2; ~~z_2 = c_2 (h-r) \delta_2,$$

\noindent where $a_2, b_2, c_2$ are positive numbers between 0 and 1 generated randomly and normalized such that

$$a_2^2 + b_2^2 + c_2^2 = 1.$$

\noindent By analogy with the first capsule, $\delta_2$ is a discrete variable that goes from $-1$ to $+1$ in steps of 0.01.

\begin{figure*}
\centering
\includegraphics[angle=0, scale=0.4]{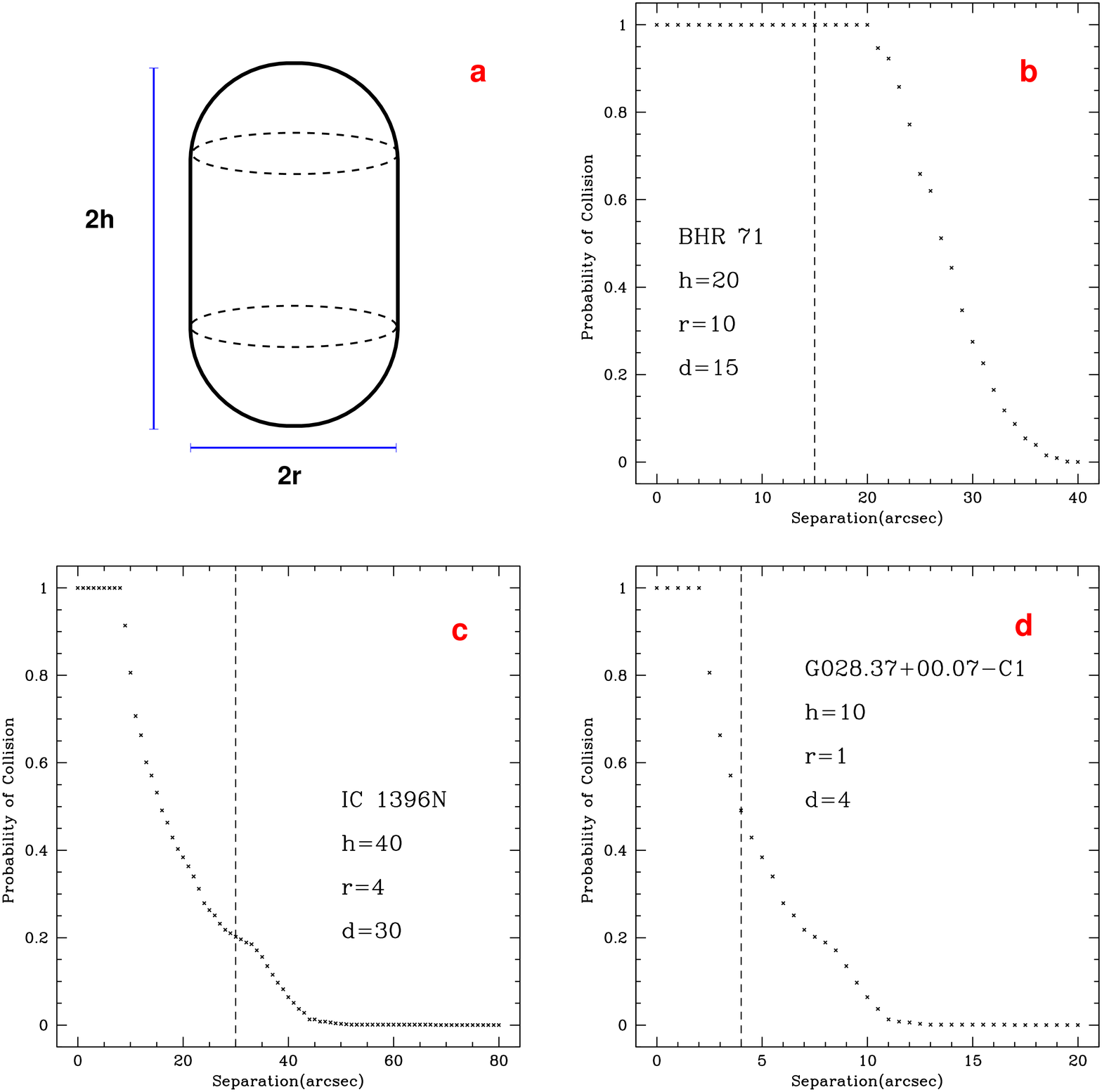}
\caption{
a) Dimensions of a circular right capsule.
b) Probability plot for the binary outflows in BHR~71 (this paper).
c) Probability plot for the binary outflows in IC 1396N \citep{bel2012}.
d) Probability plot for the binary outflows in G028.37+00.07-C1 \citep{kon2017}.
} 
\end{figure*}

For every set of $a_1, b_1, c_1, a_2, b_2, c_2$ numbers (representing a given orientation of the capsules) we calculate the separation
between the points of the axes running across the $\delta_1$ and $\delta_2$ discrete parameters. We consider that the two capsules
overlap if $s$, the separation between points, is less than twice the radius of the cylinder:

$$s < 2r.$$

For each separation we run a realization with $10^5$ sets of $a_1, b_1, c_1, a_2, b_2, c_2$ values and the probability is calculated from
these realizations. There are two limit cases. If $d < 2r$ the probability of collision is 1, while if $d > 2h$ the probability of collision is 0.

In the specific case of BHR~71, we adopt as values $r = 10"$, $h=20"$, and $d=15"$. Here we are making the assumption that the dimensions in the
plane of the sky represent the true dimensions. The resulting probability plot is shown in Figure 6b. As can be seen there, for a separation of $d=15"$,
the probability of collision is very close to 1, regardless of the orientation of the flows. This results from the fact that these flows are wide and close to each
other.

In contrast, there are cases of binary flows reported in the literature where a similar analysis indicates that a change in
the orientation of the flows could have avoided the collision. We have modeled the colliding outflows reported by \citet{bel2012,kon2017} 
with the parameters given in the Figure 6c and 6d. In these two cases the probabilities of collision are 0.2 and 0.5, respectively.
This means that a change of orientation in the flows could have avoided the collision.



\end{document}